# Spin scattering and noncollinear spin structure-induced intrinsic anomalous Hall effect in antiferromagnetic topological insulator MnBi$_2$Te$_4$

Seng Huat Lee[1,2†,] Yanglin Zhu[3†], Yu Wang[1,2], Leixin Miao[4], Timothy Pillsbury[2],

Susan Kempinger[2], David Graf[5], Nasim Alem[4], Cui-Zu Chang[1,2], Nitin Samarth[1,2] and

Zhiqiang Mao[1,2*]

[1]2D Crystal Consortium, Materials Research Institute, Pennsylvania State University, University Park, Pennsylvania 16802, USA

[2]Department of Physics, Pennsylvania State University, University Park, Pennsylvania 16802, USA

[3]Department of Physics and Engineering Physics, Tulane University, New Orleans, Louisiana 70118, USA

[4]Department of Materials Science and Engineering, Pennsylvania State University, University Park, Pennsylvania 16802, USA

[5]National High Magnetic Field Lab, Tallahassee, Florida 32310, USA

**MnBi$_2$Te$_4$ has recently been established as an intrinsic antiferromagnetic (AFM) topological insulator and predicted to be an ideal platform to realize quantum anomalous Hall (QAH) insulator and axion insulator states. We performed comprehensive studies on the structure, nontrivial surface state and magnetotransport properties of this material. Our results reveal an intrinsic anomalous Hall effect arising from a non-collinear spin structure for the magnetic field parallel to the *c*-axis. We also observed remarkable negative magnetoresistance under arbitrary field orientation below and above the Neel temperature (T$_N$), providing clear evidence for strong spin fluctuation-driven spin scattering in both the AFM and paramagnetic states. Further, we found that the nontrivial surface state opens a large gap (~85 meV) even at temperatures far above $T_N$ = 25K. These findings demonstrate that the bulk band structure of MnBi$_2$Te$_4$ is strongly coupled to the magnetic structure and that a net Berry curvature in momentum space can be created in a canted AFM state. In** **addition, our results imply that the gap opening in the surface states is intrinsic, likely caused by the strong spin fluctuations near the surface layers.**

[†]S.H. Lee and Y.L. Zhu equally contributed to this work.

*email: zim1@psu.edu

The discovery of the quantum anomalous Hall (QAH) insulators has attracted tremendous interest in the scientific community [1-7], since it can support a spin-polarized dissipation-less chiral edge states that has potential for applications in energy-efficient electronics and spintronics. The realization of the QAH effect requires a combination of magnetism and topological electronic states [2,6,7]. The QAH effect has been realized in thin films of Cr- and/or V-doped $(Bi,Sb)_2Te_3$, a topological insulator (TI) [3,4] wherein magnetism is introduced by the magnetic dopants Cr or V. In addition to magnetic doping, the magnetic order can also be introduced into a TI layer through proximity to a ferromagnetic (FM) or an antiferromagnetic (AFM) insulator layer [8-14]. To date, the fully developed QAH state has only been realized in magnetically doped TI films at a 'critical temperature' below ~1K [3,4,15,16]. This severely constrains both the exploration of fundamental physics and meaningful technological applications based on this exotic phenomenon. Therefore, seeking higher temperature QAH insulators has been a major direction in the field of topological quantum materials. One of the proposed approaches is to realize a high-temperature QAH insulator in thin films of intrinsic FM or AFM TI materials [17]. Nevertheless, despite considerable theoretical and experimental efforts, there has been little progress until the recent discovery of an intrinsic AFM TI $MnBi_2Te_4$ [18-20].

$MnBi_2Te_4$ is a tetradymite-type compound [21], a variant of a well-known TI $Bi_2Te_3$; it crystallizes in a rhombohedral structure with the space group $R\bar{3}m$, built of the stacking of Te-Bi-Te-Mn-Te-Bi-Te septuple layers (SLs) along the $c$-axis (inset of Fig. 1b). SLs are coupled through van der Waals bonding. Each SL can be viewed as being formed from the intercalation of a MnTe bilayer into a quintuple layer of $Bi_2Te_3$. This material allows for the combination of antiferromagnetism and nontrivial topological states, thus giving rise to an intrinsic AFM

TI [18-20]. Its antiferromagnetism is produced by the Mn-sub-lattice, while its nontrivial surface state is formed by inverted Bi and Te $p_z$ bands at the Γ point due to strong spin-orbital coupling. Since its AFM state is characterized by an *A*-type AFM order ($T_N$ = 25K) [20], formed by the Mn FM layers stacked antiferromagnetically along the *c*-axis and the ordered magnetic moments are aligned with the *c*-axis [20], the FM layer near the cleavage surface is anticipated to break time-reversal symmetry (TRS), thus opening a large gap (~50 or 88 meV [19,20]) in the topological surface state; this has been probed in ARPES measurements on single crystal samples [20]. Such an intrinsic AFM TI has been predicted to be an ideal platform to realize the QAH insulator; theory shows that $MnBi_2Te_4$ thin films with odd numbers of SLs support such a state [18,19,22]. Although this is yet to be verified experimentally, the successful growth of $MnBi_2Te_4$ thin films by MBE and its intrinsic AFM TI property have already been demonstrated [23]. Further, theoretical studies have also indicated that another topological quantum state, an axion insulator state [8,18] with the topological magnetoelectric effect, can be realized in $MnBi_2Te_4$ thin films with even numbers of SLs. Besides QAH and axion insulator states, $MnBi_2Te_4$ is also predicted to host the elusive ideal Weyl semimetal state with a single pair of Weyl nodes near the Fermi level [18,19] in its FM phase. The material could also generate chiral Majorana modes via interaction with a *s*-wave superconductor [24].

Since $MnBi_2Te_4$ offers opportunities to realize many variants of topological quantum states, it is imperative to understand its fundamental properties. In this *Letter*, we report a systematic study of single crystal samples of $MnBi_2Te_4$, focusing on elucidating the coupling between the bulk electronic and magnetic properties, as well as exploring how the surface states are related to magnetism. Our studies reveal clear evidence of spin fluctuation-driven spin scattering in both the AFM ordered state and in the paramagnetic (PM) state above $T_N$, suggesting that the opening of a gap in the nontrivial surface states in the PM state is likely driven by the strong spin fluctuations near the surface layers. Moreover, we find when the magnetic field is applied in the out-of-plane direction, the system evolves into a metastable canted AFM state prior to polarized to a FM state. In such a canted AFM, we observed an intrinsic anomalous Hall effect not coupled with magnetization, indicating the presence of a net Berry curvature in momentum space created by the non-collinear spin structure. These findings may accelerate experimental efforts to realize the anticipated topological quantum states in $MnBi_2Te_4$.

Platelike $MnBi_2Te_4$ single crystals (inset of Fig. 1a) were synthesized using the melt growth method, like the one reported in Ref. [20]. The excellent crystallinity of our crystal is demonstrated by the sharp (00$L$) X-ray diffraction peaks (Fig. S1). The sharp diffraction spots are seen in the electron diffraction pattern taken along the [100] zone axis (Fig. 1a), indicating the single crystallinity of the sample with no stacking faults between SLs. High resolution scanning transmission electron microscope (STEM) imaging was utilized to characterize the structure of $MnBi_2Te_4$ crystals. The atomic resolution high angle annular dark field (HAADF)-STEM image taken along the [100] zone axis clearly demonstrates the layered structure of $MnBi_2Te_4$ comprised of the stacking of Te-Bi-Te-Mn-Te-Bi-Te SLs along the $c$-axis (Fig. 1b) [21]. The previously-reported structure refinement of X-ray diffraction suggests that there exists some degree of Bi-Mn anti-site defects [20]. Figure 1c and 1d show the intensity line profile of Bi and Mn atoms indicated by red and blue arrows in the STEM image. The intensity line profiles for Bi and Mn have a uniform intensity indicating little to no intermixing of the two elements. This observation is consistent with the existing theoretical predictions suggesting that Mn-Bi intermixing is not energetically favorable [19]. Additionally, within the Mn intensity line profile, peaks with lower Mn intensity can be seen (marked by yellow arrows), indicating Mn vacancies within the atomic column. The Mn vacancies can be also clearly distinguished in the bright field (BF) and low angle annular dark field (LAADF)-STEM images (Fig S2) and the resulting strain contrast associated with them. Given that the AFL order in MnBi2Te4 is produced by the Mn-sublattice, the presence of Mn vacancies may affect its magnetism and magnetotransport properties.

A key signature of the AFM TI state in $MnBi_2Te_4$ is that its surface with preserved $S$ symmetry has a gapless Dirac cone, while the surface with broken $S$ symmetry is characterized by a Dirac cone with a large gap (50 or 88 meV [19,20]). $S$ stands for a combined symmetry, defined as $S = \Theta T_{1/2}$, where $\Theta$ represents TRS and $T_{1/2}$ denotes primitive-lattice transition symmetry [25]. The (001) surface, the cleavage plane, is a $S$-symmetry broken surface in the AFM state and thus should have a gapped Dirac cone [19]. Such a gapped Dirac cone has recently been probed in the ARPES band maps of single crystal samples [20]; the gap magnitude measured at $\Gamma$ point is ~ 70 meV at 17K, consistent with the theoretically calculated gap. Intriguingly, such a gap was found to remain open even up to room temperature for single crystal samples, inconsistent with the ARPES results on MBE-grown $MnBi_2Te_4$ thin films [23] which shows the Dirac cone of the (001) surface is gapless at temperatures above $T_N$. To see if our samples show a gapped Dirac cone state on the (001) surface, we performed ARPES measurements on our $MnBi_2Te_4$ single crystal samples and observed clear signatures of gapped surface states at both 30K and 300K (Figs. 2a and 2b). The gap magnitude is ~85 meV at room temperature, comparable to the surface state gap size reported in Ref. [20]. Such a surface state gap opening is most likely driven by spin fluctuations. We observed strong evidence of spin-scattering caused by spin fluctuations in the PM state from magnetotransport measurements, as will be discussed below.

Figure 3a shows the temperature dependence of the magnetic susceptibility measured with the field cooling (FC) and zero-field cooling (ZFC) histories and magnetic field aligned parallel ($H//c$) and perpendicular ($H\perp c$) to the $c$-axis respectively. These data demonstrate an AFM transition at ~25K with the spin-easy axis along the $c$-axis. The isothermal magnetization measurements performed at 5K for $H//c$ (Fig. 3b) reveals a remarkable spin-flop transition near 3.57 T (denoted by $H_{c1}$ hereafter). This magnetic behavior is consistent with a previous report [20]. We note that the magnetization shows a linear increase above $H_{c1}$ and the ordered moment at the maximum field (5 T) of our measurements is ~2.2 $\mu_B$/Mn, lower than the effective moment of 5.0 $\mu_B$/Mn expected for $Mn^{2+}$ ion with the half-filled $3d$ state in $MnBi_2Te_4$, indicating that the spin flop transition leads the system to a metastable canted AFM (CAFM) state prior to being polarized to a FM state. To probe the CAFM-to-FM transition, we conducted magnetic torque measurements under high magnetic fields and found the transition from the CAFM to FM phase occurs at ~7.70 T (denoted by $H_{c2}$), as shown in Fig. 4a, where both the torque and magnetization data are included for comparison and schematics of three distinct magnetic phases (i.e. AFM, CAFM and FM) in the different field ranges are also presented. The linear increase of magnetization in the CAFM phase suggests that the moment canting angle of this phase should vary with magnetic field. The metamagnetic transition does not appear for $H\perp c$. The isothermal magnetization measured at 5K for $H\perp c$ displays a linear field dependence (Fig. 3b), indicating that the AFM state is gradually polarized to an FM state without undergoing a metastable CAFM state. Our high field torque measurements (Fig. 4d) and magnetotransport measurements (Figs. 4e to 4f) indicate the full FM polarization occurs around 10.5 T.

Figures 3c and 3d present the temperature dependences of the in-plane ($\rho_{xx}$) and out-of-plane ($\rho_{zz}$) resistivity under different magnetic fields (applied along the c-axis). Both $\rho_{xx}(T)$ and $\rho_{zz}(T)$ at zero field exhibit metallic behavior above 40K with a moderate electronic anisotropy ($\rho_{zz}/\rho_{xx}$ ~ 35 at 2K) and remarkable peaks at the AFM transition temperature ($T_N$ = 25K). Such resistivity peaks near $T_N$ can be well understood in terms of spin scattering due to spin fluctuations as discussed below. In general, in AFMs or FMs, spin fluctuations always tend to intensify as the temperature approaches the ordering temperature where they reach maximum strength. For a local-moment AFM like $MnBi_2Te_4$, the intensified spin fluctuations due to spin-wave excitations near $T_N$ can greatly affect transport properties through spin-scattering if itinerant carriers interact with local moments. This results in a resistivity peak near $T_N$. Such a scenario was previously demonstrated in an AFM material $Fe_{1+y}Te$ [26]. The resistivity peak near $T_N$ observed here should follow a similar mechanism.

To further verify the involvement of the spin fluctuations driven spin scattering in transport, we systematically measured the variation of resistivity under an external magnetic field. Figures 3c and 3d show that the peaks near $T_N$ in both $\rho_{xx}(T)$ and $\rho_{zz}(T)$ are suppressed by the magnetic fields higher than $H_{c1}$. This observation is consistent with the spin-fluctuation-driven spin scattering scenario. Further, for an A-type AFM system, interlayer AFM coupling usually generates strong spin scattering in the ordered state, resulting in a high resistivity state. However, when the applied field is strong enough to overcome the Zeeman energy of the interlayer AFM coupling and push it to a forced FM state, the spin-scattering would be suppressed, resulting in a low resistivity state. Such a phenomenon is termed as a spin-valve effect and was first demonstrated in magnetic, multi-layer thin films [27,28]. The spin-valve effect can also occur between an *A*-type AFM and a CAFM state as seen in $Ca_3Ru_2O_7$ [29]. We indeed observed a similar spin-valve effect in the magnetotransport measurements of $MnBi_2Te_4$, as shown below.

We performed magnetic field sweep measurements of both $\rho_{xx}$ and $\rho_{zz}$ at various temperatures for both $H//c$ and $H \perp c$, as presented in Fig. 4. For $H//c$, $\rho_{xx}(H)$ and $\rho_{zz}(H)$ show steep decrease in response to the AFM-to-CAFM transition at $H_{c1}$ (Fig. 4b and 4c), consistent with the expected spin valve behavior, indicating the interlayer AFM ordering results in strong spin scattering. When the system is across the CAFM-to-FM transition at $H_{c2}$, $\rho_{xx}(H)$ tends to be constant, while $\rho_{zz}(H)$ displays a little hump at $H_{c2}$ and this hump shifts to lower field at higher temperatures. These observations suggest that the spin scattering due to the interlayer AFM coupling is mostly suppressed in the CAFM phase. The little hump near $H_{c2}$ observed in $\rho_{zz}(H)$ is likely due to the fact that the magnetoresistivity also contains a positive term besides a negative term that is due to the suppression of spin scattering. The positive term grows larger under high magnetic fields and becomes significant above 15 T (Fig. S3). The positive magnetoresistance (MR) term should be induced by the Lorentz effect and anisotropic Fermi surface.

The suppression of spin-scattering by magnetic field is also observed in $\rho_{xx}(H)$ and $\rho_{zz}(H)$ measurements for $H \perp c$ (Figs. 4e and 4f). Since the AFM phase is gradually polarized to a FM state in this field configuration (Fig. 4d), $\rho_{xx}(H)$ and $\rho_{zz}(H)$ show a gradual decrease with increasing field and then tend to level off near $H_{c2}$. In addition to observing spin-scattering in the AFM state, we also found spin scattering is substantial in the PM state, as evidenced by the large negative MR at $T$ = 30K for both $H//c$ and $H\perp c$ (e.g. $[\rho_{zz}(H)-\rho_{zz}(0)]/\rho_{zz}(0)$ ~ -6-9% at 12.5 T and 30K, comparable to the magnitude of negative MR seen in the AFM state in the same field range). These results strongly support that the PM state has strong spin fluctuations. Since the intralayer Mn-Te-Mn FM super-exchange interactions is much stronger than the interlayer AFM interaction, the spin fluctuations in the PM state should feature FM correlation (i.e. FM spin fluctuations), as manifested in the positive Curie-Weiss temperature ($\theta_{CW}$ ~ 5K) extracted from the susceptibility (inset of Fig. 3a). Such strong FM fluctuations might break TRS symmetry, thus resulting in the gap opening of surface states as observed in ARPES experiments (Fig. 2). The spin fluctuation driven gap opening in the nontrivial surface state in magnetic doped TIs has indeed been established in previous work [30].

Under strong spin scattering, another possible consequence is low transport mobility. This is also verified in our Hall resistivity $\rho_{xy}$ measurements. Figure 5a shows $\rho_{xy}$ as a function of magnetic field at various temperatures. From these data, we not only find the carrier mobility is indeed low, but also observed an intrinsic anomalous Hall (AH) effect in the CAFM phase. When the field is increased above $H_{c2}$, all the data taken at different temperatures collapse into a single linear line and such a linear dependence extend to our maximum measurement field 35T (inset of Fig. 5a), consistent with the single-band nature of this system. The Hall coefficient $R_0$ extracted from such a linear field dependence of $\rho_{xy}$ is – 4.76×10$^{-10}$ Ω.cm/Oe, from which the carrier mobility $\mu$ (=$R_0/\rho_{xx}$) is estimated to 79 cm$^2$/Vs at 30K, much less than the carrier mobility in bulk Bi$_2$Te$_3$ (> 800 cm$^2$/Vs) [31,32]. This verifies the above conjecture that spin scattering reduces the carrier mobility. The carrier density estimated from $R_0$ is 1.31×10$^{20}$ cm$^{-3}$, falling into the carrier density range of $Bi_2Te_3$ ($10^{17}$ ~ $10^{20}$ cm$^{-3}$) [32,33]. The negative sign of $R_0$ indicates that the carriers in our samples are electron-type, consistent with the ARPES measurement results (Fig. 2).

Next, we focus on the intrinsic AH effect observed in the CAFM phase. The $\rho_{xy}$ of this phase shows a striking non-linear field dependence, contrasted with the perfect linear dependence of the FM phase and the nearly-linear dependence of the AFM phase. To better illustrate this feature, we plot the field dependence of -$\rho_{xy}$ at T = 1.5K separately, in Fig. 5b. Apparently, the linear increase above $H_{c2}$ is from the normal Hall resistivity $\rho_{xy}^N$ (Fig. 5b). We obtain AH resistivity $\Delta\rho_{xy}$ after subtracting $\rho_{xy}^N$ (Fig. 5b). $\Delta\rho_{xy}$ can be further expressed as $\Delta\rho_{xy} = \rho_{xy}^A + \rho_{xy}^T$ [34], where $\rho_{xy}^A$ represents the AH resistivity coupled to the magnetization, whereas $\rho_{xy}^T$ stands for the AH resistivity due to the net Berry curvature in the momentum space caused by the non-collinear spin structure. $\rho_{xy}^A$ arises from the combination of the extrinsic (the skew scattering and the side jump) and intrinsic contributions and is proportional to magnetization [35]. To separate $\rho_{xy}^T$ from $\Delta\rho_{xy}$, we plot $\Delta\rho_{xy}$ as a function of magnetization $M$ in Fig. 5c (note that the magnetization data is only up to 5T). An apparent linear dependence between $\Delta\rho_{xy}$ and $M$ can be seen in the 3.5T~3.6T narrow field range where the AFM-to-CAFM transition occurs. From the linear extrapolation, we find the $\rho_{xy}^T$ contribution in the CAFM phase is large, ~6.7 μΩ.cm (Fig. 5c), comparable to that seen in the chiral antiferromagnet $Mn_3Sn$ [36]. The low-field AFM phase also shows a non-zero $\rho_{xy}^T$ though it is much smaller than that of the CAFM phase.

A large $\rho_{xy}^T$ is generally expected for an AFM state with non-collinear spin structure [37]. Symmetry breaking plus strong spin orbital coupling in such systems can lift spin degeneracy, which can lead to a net Berry curvature in momentum space, thus resulting in intrinsic AH effect. There have been several established examples for this theoretical prediction, including Mn$_3$Sn [36] and GdPtBi [34]. The large $\rho_{xy}^T$ we observed here in the CAFM phase should follow a similar mechanism, implying that the electronic band structure of MnBi$_2$Te$_4$ is strongly coupled to its magnetism, which offers an opportunity to observe new topological states tuned by a magnetic field. Theoretical studies have shown the AFM TI state can evolve into an ideal Weyl semimetal state when the system transforms into a FM phase under magnetic field. However, we did not observe any transport signatures of a Weyl state in the FM phase such as chiral anomaly. The negative MR observed in MnBie$_2$Te$_4$ is primarily from the suppression of spin scattering and appear in both the *H//I* and *H⊥I* configurations (see Figs. 4b, 4c, 4e and 4f). The absence of the chiral anomaly is probably because of the Weyl node far from the Fermi level.

In summary, the strong coupling between electronic and magnetic properties of MnBi$_2$Te$_4$ provides clear evidence of spin-scattering in both the AFM and PM states. The spin scattering of AFM state originates from the anti-parallel spin alignment of Mn ions between SLs and is mostly suppressed when the AFM state is switched to a canted AFM state for *H//c*. The spin fluctuations in the PM state also cause remarkable spin scattering and are likely to be responsible for the gap opening of surface states above $T_N$. Moreover, we find the canted AFM that exhibits an intrinsic AH effect as a result of the net Berry curvature in the momentum space induced by the non-collinear spin structure, indicating that the electronic band structure of MnBi$_2$Te$_4$ is strongly coupled with magnetism. Our findings will help us understand the interplay between the magnetism and topological surface state in this newly-discovered intrinsic AFM TI material and will facilitate the realization of the QAH insulator and axion insulator states in thin films of this material.

**Acknowledgement**

The study is based upon research conducted at The Pennsylvania State University Two-Dimensional Crystal Consortium–Materials Innovation Platform (2DCC-MIP) which is supported by NSF cooperative agreement DMR-1539916. Y.L.Z. is supported by the U.S. Department of Energy under EPSCoR grant no. DESC0012432 with additional support from the Louisiana Board of Regents. The work at the National High Magnetic Field Laboratory is supported by the NSF Cooperative Agreement No. DMR-1157490 and the State of Florida. We thank Chaoxing Liu and Jun Zhu for useful discussions.

**Figures Captions**

**Figure 1. Crystal structure of MnBi$_2$Te$_4$.**

(a) The selected area diffraction pattern taken from [100] zone axis. The sharp diffraction spots indicate no stacking faults in the measured sample. Inset: Image of a platelike MnBi$_2$Te$_4$ single crystal grown via melt growth method. (b) HAADF-STEM image of the MnBi$_2$Te$_4$ crystal taken from the [100] zone axis. Inset: Magnified image of a one-unit cell in the dashed yellow rectangle with the atoms overlaid on top to demonstrates the layered structure in the stacking order of Te-Bi-Te-Mn-Te-Bi-Te septuple layers along the *c*-axis. The red and blue arrows indicate the Mn and Bi atomic columns used for the image intensity analysis. The Mn vacancies are visible at the columns indicated by the yellow arrows. The scale bar is 1nm. (c) The intensity profile of Mn atomic columns indicated by the red arrows in (b). The drop of the intensity in the first few peaks indicates the existence of the Mn vacancies in the crystal. (d) The intensity profile of Bi atomic columns indicated by the blue arrows in (b).

**Figure 2 ARPES measurements on MnBi$_2$Te$_4$ single crystals.**

ARPES band maps of MnBi$_2$Te$_4$ single crystal along the K $-$ Γ $-$ K direction at (a) room temperature, 300K and (b) at 30K, which is near to the $T_N$ (= 25K). Both ARPES data show a gap opening in the surface Dirac cone, ~85 meV and 115 meV at 300K and at 30K, respectively.

**Figure 3. Magnetic properties of MnBi$_2$Te$_4$ and its influence on transport signatures.**

(a) Field cooled (FC) and zero-field cooled (ZFC) temperature dependences of magnetic susceptibility $\chi$ measured at 1 T for the crystals aligned with the magnetic field parallel ($H//c$) and perpendicular ($H\perp c$) to the $c$-axis, respectively. The susceptibility data shows an AFM transition with $T_N$ = 25K. Inset: Modified Curie-Weiss plot with $\chi_o$ = -0.007 emu/mol Oe. The fit of susceptibility data in the 100-300 K range by the modified Curie-Weiss law yields $\theta_{CW}$ = 5.0(03)K. (b) Isothermal magnetization $M$ for the crystal aligned with magnetic field parallel ($H//c$) and perpendicular ($H\perp c$) to the $c$-axis at 5K. The system transits from an AFM to a CAFM state via a spin-flop transition at $H_{c1}$ = 3.57 T for $H//c$, while for $H\perp c$ the magnetization increases linearly with field, indicating gradual polarization to a FM state. (c) and (d), temperature dependences of the in-plane resistivity $\rho_{xx}$ and out-of-plane resistivity $\rho_{zz}$ under various magnetic fields (applied along the out-of-plane direction).

**Figure 4. Field-driven magnetic transitions and magnetotransport properties of MnBi$_2$Te$_4$.**

(a), (b), (c) Field-dependent torque $\tau$, magnetization $M$, in-plane resistivity $\rho_{xx}$ and out-of-plane resistivity $\rho_{zz}$ for field parallel to $c$-axis ($H//c$) at various temperatures. The schematics of AFM, CAFM and FM in the different field ranges are illustrated in (a). The CAFM angle is ~56° relative to the $c$-axis, estimated from the averaged moment between $H_{c1}$ and $H_{c2}$. $\tau$, $M$, $\rho_{xx}$, and $\rho_{zz}$ for field perpendicular to the $c$-axis ($H\perp c$) are also shown in (d)-(f). For $H\perp c$, the spin gradually transits from the AFM to FM and all spin are fully polarized at 10.5T. $c'$ in the torque measurements (a & d panels) is defined as the direction with a few degrees offset from the $c$-axis. The schematics of the magnetotransport experiments setup are shown in (b), (c), (e) and (f).

**Figure 5. Anomalous Hall (AH) effect of MnBi$_2$Te$_4$**

(a) Hall resistivity $\rho_{xy}$ as a function of magnetic field at various temperatures. The schematic illustrates the setup of the Hall resistivity measurements. Inset: $\rho_{xy}$ vs. $H$ in the 0-35 T field range at 1.5 K and 30K. (b) -$\rho_{xy}$ plotted against $H$ for 1.5K. $\Delta\rho_{xy}$ represents the anomalous Hall resistivity, obtained by subtracting the normal Hall resistivity $\rho_{xy}^N$ (=$R_0H$) caused by the Lorentz effect from the measured $\rho_{xy}$. (c) $\Delta\rho_{xy}$ as a function of the magnetization measured at 5K. $\rho_{xy}^T$ stands for the Hall resistivity not coupled with magnetization.

**Figure 1**

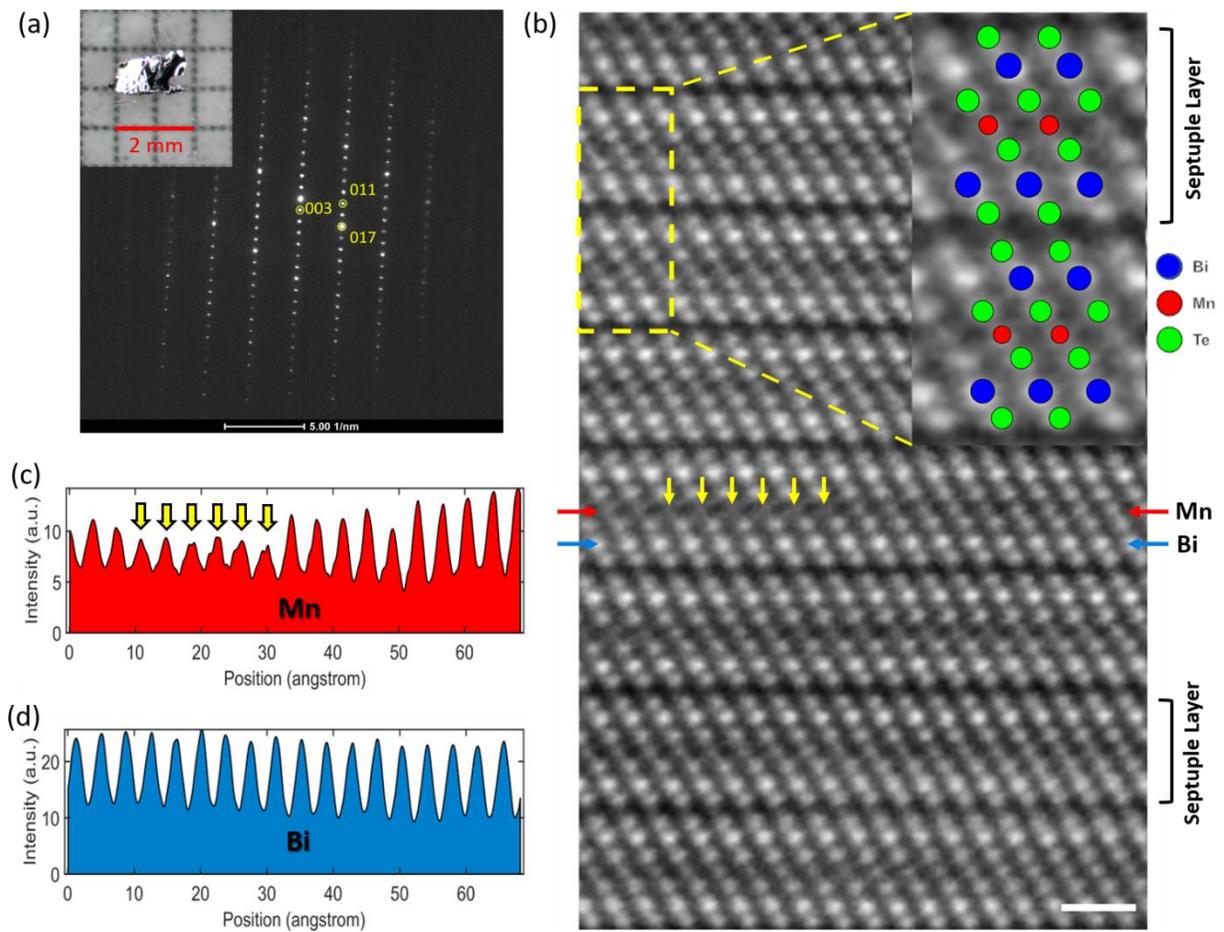

**Figure 2**

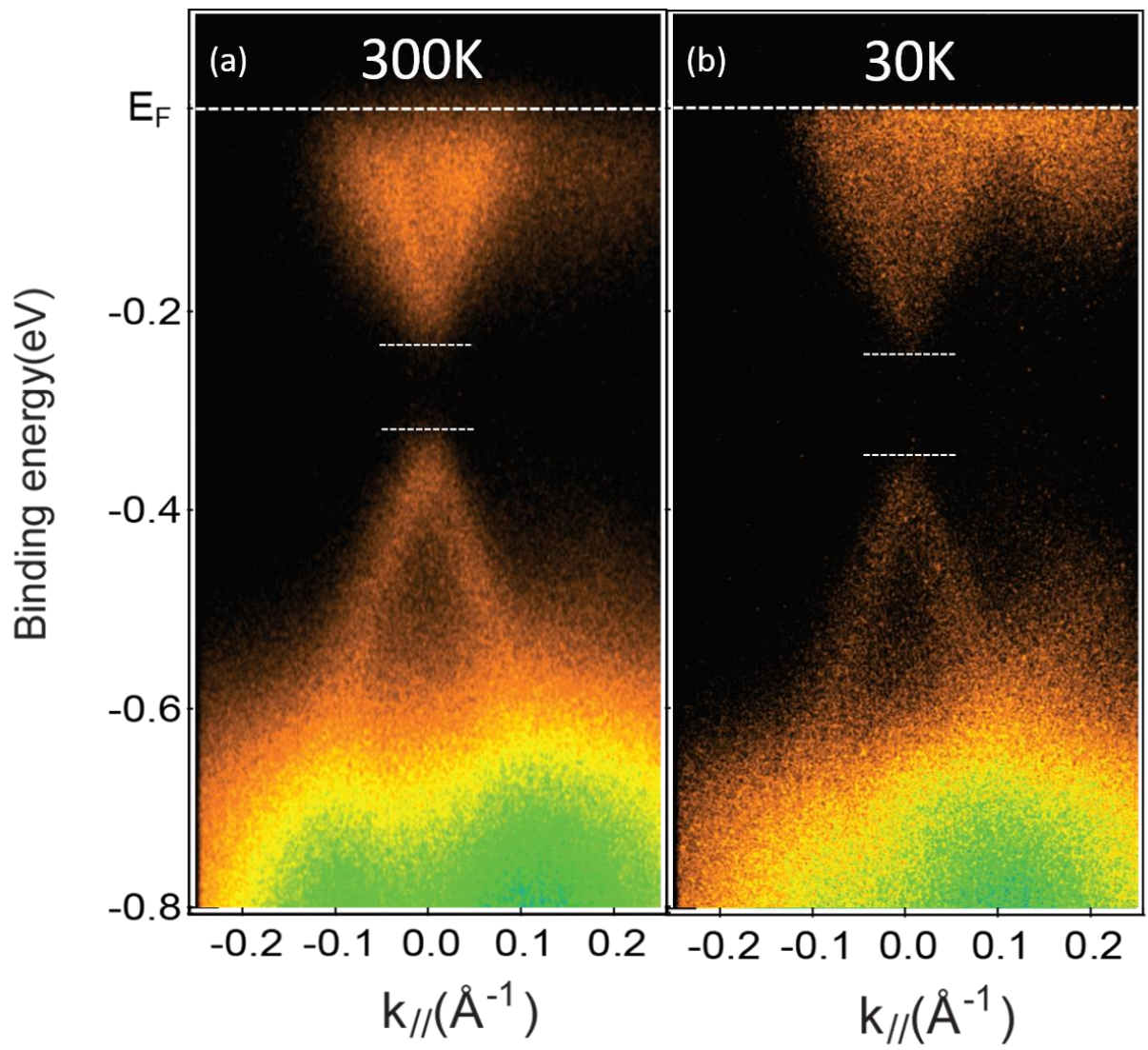

**Figure 3**

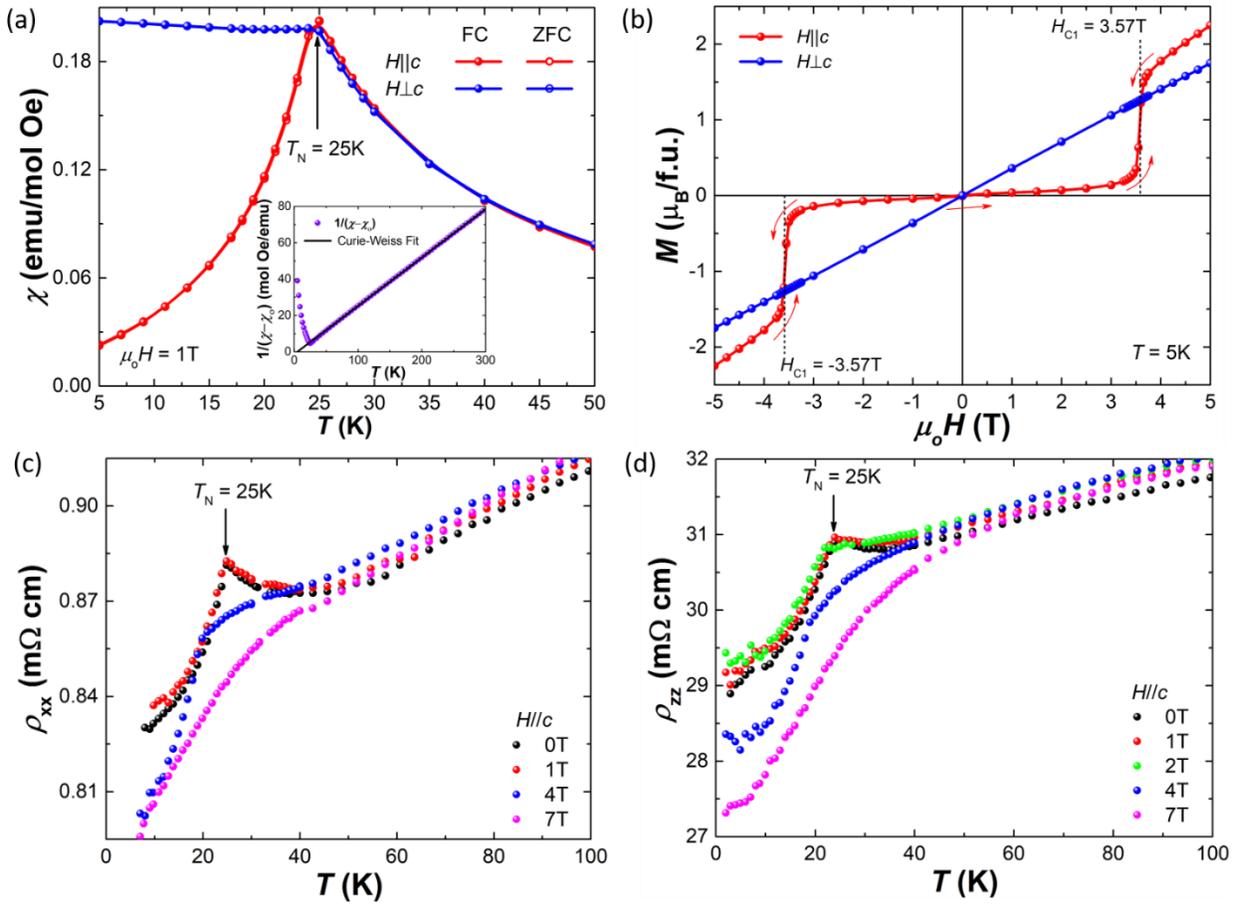

**Figure 4**

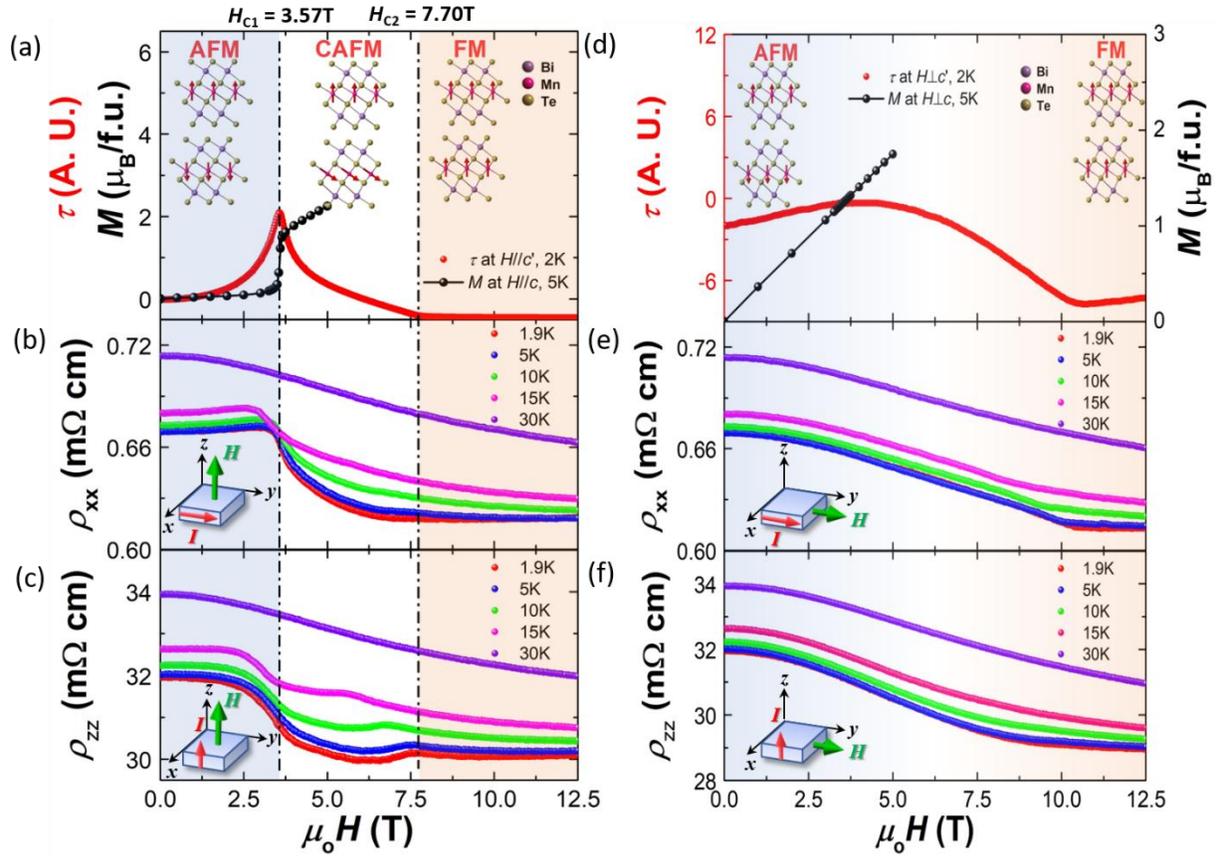

**Figure 5**

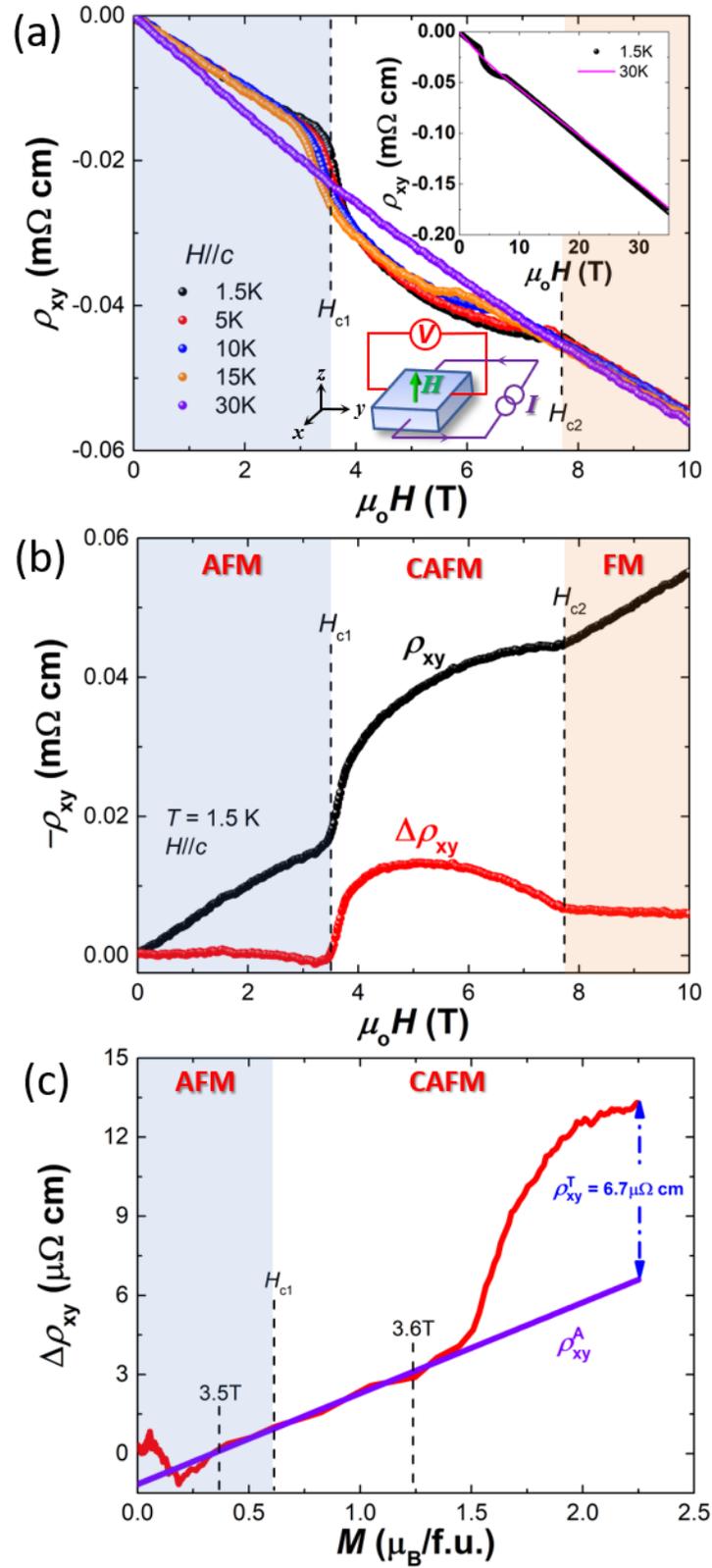

# Supplementary Information

## S1 Crystal Growth

High-quality single crystals of MnBi$_2$Te$_4$ were grown by a melt growth method. A stoichiometric mixture of high purity MnTe (99.9%) and Bi$_2$Te$_3$ (99.999%) was sealed in an evacuated, carbon-coated quartz tube jacked by another evacuated quartz tube. The carbon-coated wall inside the quartz tube was finished by using the thermal decomposition of acetone and it was used to avoid the direct reaction of manganese with the quartz tube. The prepared ampoule was heated up to 1000 °C before ramp-down to 610 °C with the ramp-rate of 3 °C/hr for the annealing process. The crystalline ingot was quenched in water after two days at the annealing temperature. Platelike single crystals were easily obtained by cleaving along the basal plane from the resultant ingot. Figure S1 shows the sharp (00*L*) X-ray diffraction peaks matched-well with the ICDD database PDF card 04-020-8214, indicating the excellent crystallinity of the MnBi$_2$Te$_4$ crystal. Additionally, we find MnBi$_2$Te$_4$ can be easily intergrown with Bi$_2$Te$_3$ and/or MnBi$_6$Te$_9$ and performed careful XRD screening for all the samples we have prepared. Only pure MnBi$_2$Te$_4$ single crystals were used for STEM, ARPES, magnetic, resistivity, magnetoresistance, torque and Hall measurements.

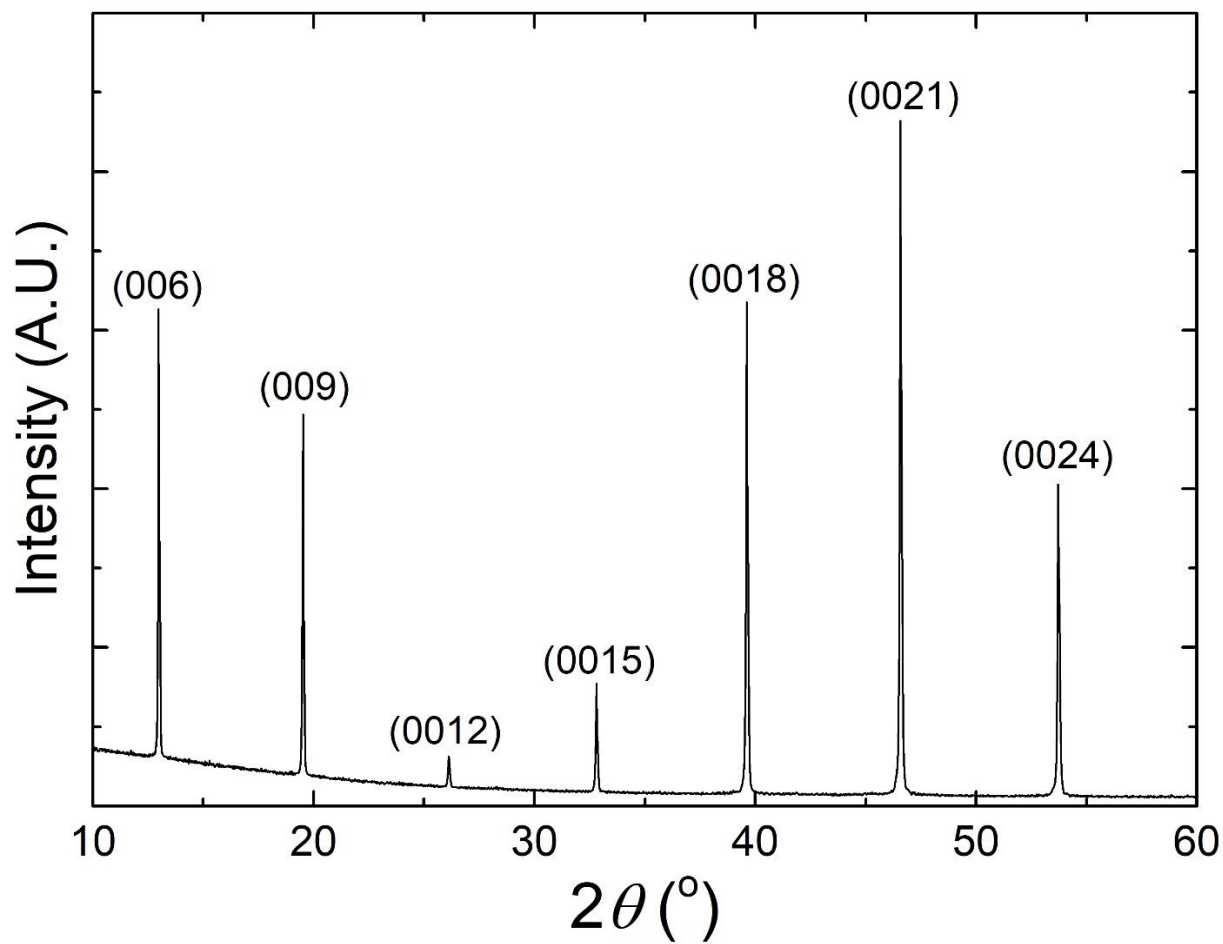

**Figure S1.** XRD pattern of a high-quality MnBi$_2$Te$_4$ crystal All peaks are matched-well with the (00*L*) peaks of ICDD PDF card 04-020-8214, demonstrating the excellent crystallinity of our sample.

## S2 STEM Analysis

The electron transparent TEM sample was prepared using an FEI Helios NanoLab Dual-Beam Focused Ion Beam system. It was thinned down using 30 kV and subsequently 5 kV Ga ion beam, and was further polished at 2kV to remove the redeposition and amorphous damage layer. The selected area electron diffraction pattern (SAEDP) was taken on the FEI Talos S/TEM at an operating voltage of 200 kV. The STEM imaging was performed on a spherical aberration corrected FEI Titan[3] S/TEM at an operating voltage of 200 kV with a probe convergence semi-angle of 30 mrad and the point resolution of 80 pm.

The HAADF detector collects the electrons that are scattered to the outer angles, and the collected signals are roughly proportional to the atomic number $Z^2$. It allows us to identify the chemical identity of all the atoms in the lattice. Since the intensity collected is formed from all the atoms in one atomic column, the Mn vacancies result in a reduction in the intensity of Mn atomic columns in HAADF-STEM image as shown in Figure S2a. The simultaneously acquired BF-STEM image in Figure S2b shows the missing Mn atomic columns at the same positions. Combining with the HAADF image, it provides more proof to the existence of the Mn vacancies.

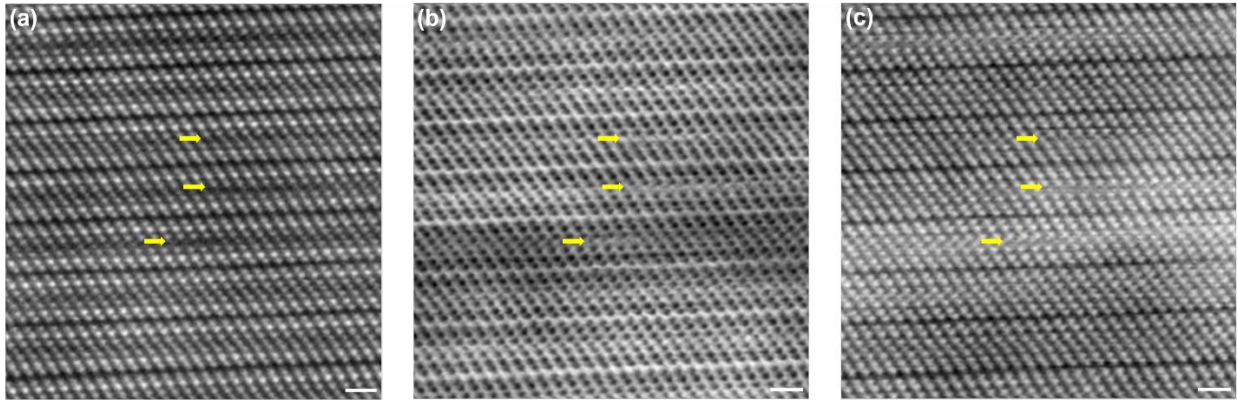

**Figure S2.** Simultaneously acquired (a) HAADF-STEM, (b) BF-STEM, and (c) LAADF-STEM from the [100] axis of the MnBi$_2$Te$_4$ crystal. The locations of the Mn vacancies are indicated by the yellow arrows.

Moreover, the septuple layers in LAADF-STEM image in S2c exhibit bright contrast at the positions where the vacancies sit. In the LAADF-STEM, the more coherently scattered electrons with lower scattering angles are collected compared to the HAADF detector. Due to the diffraction contrast present at lower scattering angles, the strain associated with the defects can be observed leading to a variation in the contrast.

## S3 Magnetotransport Measurements up to 35 T

Figure S3 shows the magnetotransport measurements for both $H//c$ and $H\perp c$ up to 35T at various temperatures. These experiments were conducted at the National High Magnetic Field Laboratory.

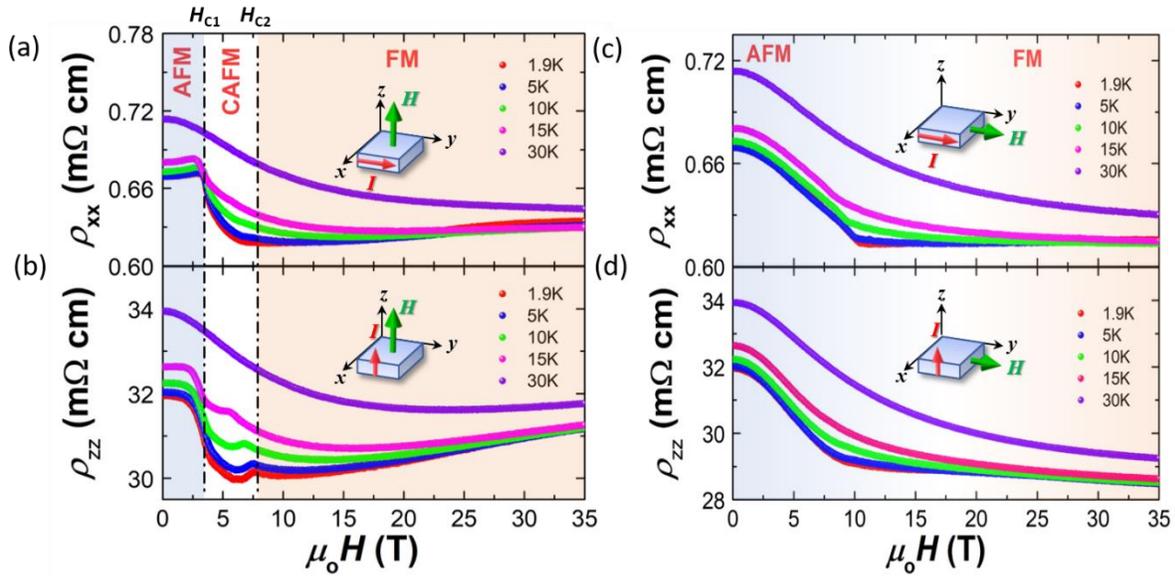

**Figure S3.** (a) and (c) show the in-plane resistivity $\rho_{xx}$ for $H//c$ and $H\perp c$, respectively. (b) and (d) show the out-of-plane resistivity $\rho_{zz}$ for $H//c$ and $H\perp c$, respectively. The schematics illustrate the setup of the magnetotransport experiments.